\documentstyle[12pt,fleqn]{article}
\begin{document}
\baselineskip=8mm

\begin{center}
{\large\bf Environment and Energy Injection Effects in GRB Afterglows}

\vspace{5.0mm}
Z. G. Dai and T. Lu

{\em Department of Astronomy, Nanjing University, Nanjing 210093, China}

\end{center}

\begin{center}
ABSTRACT
\end{center}

In a recent paper (Dai \& Lu 1999), we have
proposed a simple model in which the steepening in the light curve of
the R-band afterglow of the gamma-ray burst (GRB) 990123 is caused
by the adiabatic shock which has evolved from an ultrarelativistic phase
to a nonrelativistic phase in a dense medium. We find that such a model
is quite consistent with observations if the medium density is about
$3\times 10^6\,{\rm cm}^{-3}$. Here we discuss this model in more details.
In particular, we investigate the effects of synchrotron self absorption
and energy injection. A shock in a dense medium
becomes nonrelativistic rapidly after a short relativistic phase.
The afterglow from the shock at the nonrelativistic stage decays more
rapidly than at the relativistic stage. Since some models
for GRB energy sources predict that a strongly magnetic millisecond pulsar
may be born during the formation of GRB, we discuss the effect of such a
pulsar on the evolution of the nonrelativistic shock through magnetic dipole
radiation. We find that after the energy which the shock obtains from
the pulsar is much more than the initial energy of the shock,
the afterglow decay will flatten significantly. When the pulsar energy 
input effect disappears, the decay will steepen again. These features are
in excellent agreement with the afterglows of GRB 980519, GRB 990510
and GRB 980326. Furthermore, our model fits very well all the observational
data of GRB 980519 including the last two detections.

\vspace{2mm}
\noindent
{\em Subject headings:} gamma-rays: bursts -- stars: pulsars -- shock waves

\newpage

\begin{center}
1. INTRODUCTION
\end{center}

In the standard afterglow shock model (for a review see Piran 1999),
a gamma-ray burst (GRB) afterglow is usually believed to be produced
by synchrotron radiation or inverse Compton scattering in
an ultrarelativistic shock wave expanding in a homogeneous medium.
As more and more ambient matter is swept up, the shock gradually
decelerates while the emission from such a shock fades down, dominating
at the beginning in X-rays and progressively at optical to radio energy
band. In this model, there are two limiting cases (adiabatic and highly
radiative) for hydrodynamical evolution of a relativistic shock.
These cases have been well studied both analytically 
(e.g., M\'esz\'aros \& Rees 1997; Wijers, Rees \& M\'esz\'aros 1997;
Waxman 1997a, b; Reichart 1997; Sari 1997; Vietri 1997; Katz \& Piran 1997;
Sari, Piran \& Narayan 1998; etc) and numerically (e.g., Panaitescu,
M\'esz\'aros \& Rees 1998; Huang et al. 1998; Huang, Dai \& Lu 1998).
A partially radiative (intermediate) case has been investigated
(Chiang \& Dermer 1998; Cohen, Piran \& Sari 1998;
Dai, Huang \& Lu 1999; Huang, Dai \& Lu 1999a).
All the studies are based on the following basic assumptions:
(1) the total energy of the shock is released impulsively before its
formation; (2) the medium swept up by the shock is homogeneous
and its density ($n$) is the one of the interstellar medium
$\sim 1\,{\rm cm}^{-3}$; and (3) the electron and magnetic field energy
fractions of the shocked medium and the index ($p$) in the accelerated
electrons' power-law distribution are constant during the whole evolution
stage. The standard model is successful at explaining the overall features
of late afterglows of some bursts such as GRB 970508:
the light curves behave according to a single unbroken power law with
decay index of $\alpha \sim -1$ as long as the observations continued
(Zharikov et al. 1998).

Each of these assumptions has been varied to discuss why some observed
afterglows deviate from that expected by the standard afterglow model.
For example, the R-band light curve of GRB 970508 afterglow peaks
around two days after the burst, and there is a rather rapid rise before
the peak which is followed by a long power-law decay. There are two models
explaining this special feature: (i) Rees and M\'esz\'aros (1998)
envisioned that a postburst fireball may contain shells with a continuous
distribution of Lorentz factors. As the external forward shock sweeps up
ambient matter and decelerates, internal shells will catch up with
the shock and supply energy into it. A detailed calculation shows that
this model can explain well this special feature (Panaitescu, M\'esz\'aros
and Rees 1998). (ii) Dai \& Lu (1998a) considered continuous
energy injection from a strongly magnetized millisecond pulsar into
the shock through magnetic dipole radiation. This model can also
account for well the observations. It is very clear that these models
don't use basic assumption (1).

There are several models in the literature that discuss the effect of
inhomogeneous media on afterglows (Dai \& Lu 1998b; M\'esz\'aros, Rees 
\& Wijers 1998; Vietri 1997), dropping the second assumption.
Generally, an $n\propto r^{-k}$ ($k>0$) medium is expected to steepen
an afterglow's temporal decay. Recently, Chevalier \& Li (1999a, b) found
that a Wolf-Rayet star wind likely leads to an $n\propto r^{-2}$ medium,
and thus if GRB 980519 resulted from the explosion of such a massive star,
subsequent evolution of a relativistic shock in this medium is 
consistent with the steep decay in the R-band light curve of the afterglow
from this burst. Another way of dropping the second assumption is
that the density of an ambient medium is invoked to be as high as
$n\sim 10^6\,{\rm cm}^{-3}$. Recent observations show that the temporal
decay of the R-band afterglow of GRB 990123 steepened about 2.5 days
after this burst (Kulkarni et al. 1999; Castro-Tirado et al. 1999;
Fruchter et al. 1999). Dai \& Lu (1999) (hereafter DL99) proposed
a plausible model in which a shock expanding in a dense medium
has evolved from a relativistic phase to a nonrelativistic phase.
They found that this model fits well the observational data if the medium
density is about $3\times 10^6\,{\rm cm}^{-3}$. They further suggested
that such a medium could be a supernova or supranova or hypernova ejecta.
Of course, the steepening in the light curves of the afterglows of
these two bursts may be due to lateral spreading of a jet, as analyzed
by Rhoads (1999) and Sari, Piran \& Halpern (1999). However,
numerical studies of Panaitescu \& M\'esz\'aros (1999), Moderski, Sikora
\& Bulik (1999), Huang et al. (1999b) and Huang, Dai \& Lu (2000) show 
that the break of the light curve is weaker and much smoother than the one 
analytically predicted when the light travel effects related to the lateral
size of 
the jet and a realistic expression of the lateral expansion speed are taken 
into account.

In basic assumption (3), the electron and magnetic field energy fractions
of the shocked medium may not be varied during whole evolution,
as argued by Wang, Dai \& Lu (1999a), who analyzed all the observational
data including both the prompt optical flash and the afterglow of
GRB 990123. However, the assumption that $p$ is constant might be
inconsistent with the early afterglow from GRB 970508
(Djorgovski et al. 1997).

In this paper, we discuss the model proposed by DL99 in more details, by 
taking into account both the synchrotron self-absorption effect in the
shocked 
medium and the energy injection effect of Dai \& Lu (1998a, c). Therefore,
our present analysis, in fact, relaxes assumptions (1) and (2). So far
the bursts whose afterglow decay steepens include GRB 980519, GRB 980326
and GRB 990510 besides GRB 990123. In particular, for the former
two of these bursts, optical observations several days later are far above
a power law decline, implying possible energy injection at such a late
stage. In section 2, we analyze the spectrum and light curve of radiation
from a shock expanding in a dense medium. In section 3, we compare our
model with observations related to GRB 980519 and infer all intrinsic
parameters and the redshift of this burst. We discuss properties of
GRB 990510 and GRB 980326 in section 4, and in the final section we give
a discussion and conclusion.

\begin{center}
2. SHOCK EVOLUTION

2.1. {\em Relativistic Phase}
\end{center}

For simplicity, we assume that a relativistic shock expanding in a dense
medium is adiabatic. The evolution of a partially radiative shock depends
on both the efficiency with which the shock transfers its bulk kinetic
energy to electrons and magnetic fields and on the efficiency with which
the electrons radiate their energy (Dai, Huang \& Lu 1999). Here we don't
consider such a shock. The Blandford-McKee (1976) self similar solution
gives the Lorentz factor of an adiabatic relativistic shock,
\begin{equation}
\gamma=\frac{1}{4}\left[ \frac{17E_0(1+z)^3}{\pi nm_pc^5t_\oplus^3}
                  \right]^{1/8}
      =1E_{52}^{1/8}n_5^{-1/8}t_\oplus^{-3/8}[(1+z)/2]^{3/8},
\end{equation}
where $E_0=E_{52}\times 10^{52}{\rm ergs}$ is the total isotropic energy,
$n_{5}=n/10^5\,{\rm cm}^{-3}$, $t_\oplus$ is the observer's time
since the gamma-ray trigger in units of 1 day, $z$ is the the redshift
of the source generating this shock, and $m_p$ is the proton mass.
We assume $\gamma=1$ when $t_\oplus=t_b$. This implies
\begin{equation}
n_5=E_{52}t_b^{-3}[(1+z)/2]^3.
\end{equation}
For $t_\oplus > t_b$, the shock will be in a nonrelativistic phase. In the
following we will see different spectra and light curves from this shock
before and after the time $t_b$. We first analyze the relativistic case.

As usual, only synchrotron radiation from the shock is considered.
To analyze the spectrum and light curve, one needs to know three
crucial frequencies: the synchrotron peak frequency ($\nu_m$), the cooling
frequency ($\nu_c$), and the self-absorption frequency ($\nu_a$).
We assume a power law distribution of the electrons accelerated
by the shock: $dn'_e/d\gamma_e\propto \gamma_e^{-p}$ for $\gamma_e
\ge\gamma_{em}$, where $\gamma_e$ is the electron Lorentz factor and
$\gamma_{em}=610\epsilon_e\gamma$ is the minimum Lorentz factor.
We further assume that $\epsilon_e$ and $\epsilon_B$ are the electron
and magnetic energy density fractions of the shocked medium respectively.
The $\nu_m$ is the characteristic synchrotron frequency of an electron
with Lorentz factor of $\gamma_{em}$, while the $\nu_c$ is the characteristic
synchrotron frequency of an electron which cools on the dynamical age of
the shock. According to Sari et al. (1998), therefore, these two frequencies,
measured in the observer's frame, can be written as
\begin{equation}
\nu_m=\frac{\gamma\gamma_{em}^2}{1+z}\frac{eB'}{2\pi m_ec}
     =7.0\times 10^8(\epsilon_e/0.1)^2\epsilon_{B,-6}^{1/2}
      E_{52}^{1/2}t_\oplus^{-3/2}[(1+z)/2]^{1/2}\,\,{\rm Hz},
\end{equation}
\begin{equation}
\nu_c = \frac{18\pi em_ec(1+z)}{\sigma_T^2B'^3\gamma t_\oplus^2}
      = 2.2\times 10^{17}\epsilon_{B,-6}^{-3/2}E_{52}^{-1/2}
        n_5^{-1}t_\oplus^{-1/2}[(1+z)/2]^{-1/2}\,\,{\rm Hz},
\end{equation}
where $\epsilon_{B,-6}=\epsilon_B/10^{-6}$, $B'=(32\pi \epsilon_B
\gamma^2nm_pc^2)^{1/2}$ is the internal magnetic field strength of
the shocked medium and $\sigma_T$ is the Thompson scattering cross section.
The self-absorption frequency has been estimated to be
\begin{equation}
\nu_a=6.0\times 10^{11}(\epsilon_e/0.1)^{-1}\epsilon_{B,-6}^{1/5}E_{52}^{1/5}
      n_5^{3/5}[(1+z)/2]^{-1}\,\,{\rm Hz},
\end{equation}
where $p=2.8$ has been used (Granot, Piran \& Sari 1998; Wijers \& Galama
1998). This estimate is valid only for $\nu_a < \nu_m$. Since $\nu_m$ 
decreases as $\propto t_\oplus^{-3/2}$ and $\nu_a$ is time invariant,
we define a time $t_a$ based on $\nu_m(t_a)=\nu_a$:
\begin{equation}
t_a=0.01(\epsilon_e/0.1)^2\epsilon_{B,-6}^{1/5}E_{52}^{1/5}n_5^{-2/5}
    [(1+z)/2]\,\,{\rm days}.
\end{equation}
When $\nu_a > \nu_m$, the self-absorption coefficient $\alpha_\nu\propto
\gamma\gamma_{em}^{p-1}B'^{(p+2)/2}\nu^{-(p+4)/2}$ (Rybicki \& Lightman 1979)
and the width of the shock $\Delta r\approx r/\gamma\propto \gamma t_\oplus$,
so, based on the optical depth $\tau=\alpha_\nu\Delta r=0.35$, we find
that the self-absorption frequency decays as $\nu_a\propto
t_\oplus^{-(3p+2)/2(p+4)}$. Thus we have
\begin{equation}
\nu_a=6.0\times 10^{11}(\epsilon_e/0.1)^{-1}\epsilon_{B,-6}^{1/5}E_{52}^{1/5}
      n_5^{3/5}[(1+z)/2]^{-1}(t_\oplus/t_a)^{-\frac{3p+2}{2(p+4)}}\,\,{\rm
Hz},
\end{equation}

The observed synchrotron radiation peak flux can be obtained by
\begin{equation}
F_{\nu_m}=\frac{N_e\gamma P'_{\nu_m}(1+z)}{4\pi D_L^2}
         =10\epsilon_{B,-6}^{1/2}E_{52}n_5^{1/2}
         \left(\frac{\sqrt{1+z}-1}{\sqrt{2}-1}\right)^{-2}\,\,m{\rm Jy},
\end{equation}
where $N_e$ is the total number of swept-up electrons and $P'_{\nu_m}
=m_ec^2\sigma_TB'/(3e)$ is the radiated power per electron per unit
frequency in the frame comoving with the shocked medium. 
For a flat universe with $H_0=65\,{\rm km}\,{\rm s}^{-1}\,{\rm Mpc}^{-1}$,
the distance to the source $D_L=2c/H_0(1+z-\sqrt{1+z})$.

After having the peak flux and three break frequencies, we can write
the spectrum and light curve of synchrotron radiation. For high frequency
$\nu>\nu_{am}\equiv {\rm max}(\nu_a,\nu_m)$, we find
\begin{equation}
F_\nu=\left \{
       \begin{array}{ll}
         (\nu/\nu_m)^{-(p-1)/2}F_{\nu_m}\propto \nu^{-(p-1)/2}
              t_\oplus^{3(1-p)/4} & {\rm if}\,\, \nu_{am}<\nu<\nu_c \\
         (\nu_c/\nu_m)^{-(p-1)/2}(\nu/\nu_c)^{-p/2}F_{\nu_m}\propto
              \nu^{-p/2}t_\oplus^{(2-3p)/4} & {\rm if}\,\, \nu>\nu_c.
        \end{array}
       \right.
\end{equation}
If $p\approx 2.8$, then the temporal decay index $\alpha=3(1-p)/4
\approx -1.35$ for emission from slow-cooling electrons or $\alpha=
(2-3p)/4\approx -1.6$ for emission from fast-cooling electrons.
In addition, the low-frequency ($\nu<\nu_{am}$) radiation should be
discussed in two cases: (i) for $\nu_a<\nu_m$, the spectrum and light
curve can be written
\begin{equation}
F_\nu=\left \{
       \begin{array}{ll}
         (\nu_a/\nu_m)^{1/3}(\nu/\nu_a)^2 F_{\nu_m}\propto
              \nu^2 t_\oplus^{1/2} & {\rm if}\,\, \nu<\nu_a \\
         (\nu/\nu_m)^{1/3}F_{\nu_m}\propto \nu^{1/3}
              t_\oplus^{1/2} & {\rm if}\,\, \nu_a<\nu<\nu_m;
        \end{array}
       \right.
\end{equation}
(ii) for $\nu_a>\nu_m$, we can obtain the spectrum and light curve,
\begin{equation}
F_\nu=(\nu_a/\nu_m)^{-(p-1)/2}(\nu/\nu_a)^{5/2} F_{\nu_m}\propto
              \nu^{5/2} t_\oplus^{5/4}.
\end{equation}
These equations show that the flux of the low-frequency radiation
increases with time.

\begin{center}
2.2. {\em Nonrelativistic Phase}
\end{center}

After sweeping up sufficient ambient matter, the shock will eventually go
into a nonrelativistic phase, viz., $t_\oplus > t_b$. In the following
we analyze the spectrum and light curve of the synchrotron radiation
from such a shock, by assuming $\nu_a>\nu_m$.

\begin{center}
2.2.1. {\em Without Any Energy Injection}
\end{center}

We first consider the widely-studied case without any energy
to be input into the shock after the GRB.
In this case, the shock's velocity $v\propto t_\oplus^{-3/5}$ and
its radius $r\approx vt_\oplus \propto t_\oplus^{2/5}$. According to DL99,
thus, the synchrotron peak frequency, the cooling frequency,
the self-absorption frequency, and the peak flux are derived as
\begin{equation}
         \nu_m=7.0\times 10^8(\epsilon_e/0.1)^2\epsilon_{B,-6}^{1/2}
               E_{52}^{1/2}t_b^{-3/2}[(1+z)/2]^{1/2}
               (t_\oplus/t_b)^{-3}\,\,{\rm Hz},
\end{equation}
\begin{equation}
         \nu_c=2.2\times 10^{17}\epsilon_{B,-6}^{-3/2}E_{52}^{-1/2}
               n_5^{-1}t_b^{-1/2}[(1+z)/2]^{-1/2}(t_\oplus/t_b)^{-1/5}
               \,\,{\rm Hz},
\end{equation}
\begin{equation}
         \nu_a=6.0\times 10^{11}(\epsilon_e/0.1)^{-1}\epsilon_{B,-6}^{1/5}
               E_{52}^{1/5}n_5^{3/5}[(1+z)/2]^{-1}(t_b/t_a)^{-\frac{3p+2}
               {2(p+4)}}(t_\oplus/t_b)^{-\frac{3p-2}{p+4}}\,\,{\rm Hz},
\end{equation}
and
\begin{equation}
F_{\nu_m}=10\epsilon_{B,-6}^{1/2}E_{52}n_5^{1/2}\left(\frac{t_\oplus}
          {t_b}\right)^{3/5}\left(\frac{\sqrt{1+z}-1}{\sqrt{2}-1}
          \right)^{-2}\,\,m{\rm Jy}.
\end{equation}
Based on these equations, we further derive the spectrum and light curve,
\begin{equation}
F_\nu=\left \{
       \begin{array}{lll}
         (\nu_a/\nu_m)^{-(p-1)/2}(\nu/\nu_a)^{5/2}F_{\nu_m}
         \propto \nu^{5/2}t_\oplus^{11/10} & {\rm if}\,\, \nu<\nu_a \\
         (\nu/\nu_m)^{-(p-1)/2}F_{\nu_m} \propto \nu^{-(p-1)/2}
         t_\oplus^{(21-15p)/10} & {\rm if}\,\, \nu_a< \nu <\nu_c \\
         (\nu_c/\nu_m)^{-(p-1)/2}(\nu/\nu_c)^{-p/2}F_{\nu_m}\propto
              \nu^{-p/2}t_\oplus^{(4-3p)/2} & {\rm if}\,\, \nu>\nu_c.
        \end{array}
       \right.
\end{equation}
We easily see that for high-frequency radiation the temporal decay index
$\alpha=(21-15p)/10$ for emission from slow-cooling electrons or
$\alpha=(4-3p)/2$ for emission from fast-cooling electrons.
If $p\approx 2.8$, then $\alpha\approx -2.1$ or $-2.2$. Comparing
this with the relativistic result, we conclude that the afterglow decay
steepens at the nonrelativistic stage.

\begin{center}
2.2.2. {\em With Energy Injection from Pulsars}
\end{center}

Some models for GRB energy sources (for a brief review see
Dai \& Lu 1998c) predict that during the formation of
an ultrarelativistic fireball required by GRB, a strongly magnetized
millisecond pulsar will be born. If so, the pulsar will continuously input
its rotational energy into the forward shock of the postburst fireball
through magnetic dipole radiation because electromagnetic waves radiated
by the pulsar will be absorbed in the shocked medium (Dai \& Lu 1998a, c). 
Since an initially ultrarelativistic shock discussed in this paper
rapidly becomes nonrelativistic in a dense medium, we next investigate
the evolution of a nonrelativistic adiabatic shock with energy injection
from a pulsar. The total energy of the shock is the sum of the initial
energy and the energy which the shock has obtained from the pulsar:
\begin{equation}
E_0+\int_0^{t_\oplus}Ldt_\oplus =E_{\rm tot} \propto v^2r^3,
\end{equation}
where $L$ is the stellar spindown power $\propto (1+t_\oplus/T)^{-2}$
($T$ is the initial spindown time scale).
The term on the right-hand side is consistent
with the Sedov solution. Please note that $L$ can be thought of as
a constant for $t_\oplus<T$, while $L$ decays as $\propto t_\oplus^{-2}$
for $t_\oplus\gg T$. Because of this feature, we easily integrate the
second term on the left-hand side of equation (17). We now define a time
at which the shock has obtained energy $\sim E_0$ from the pulsar,
$t_c=E_0/L$, and assume $t_c\ll T$. We next analyze the evolution of the
afterglow from such a shock at three stages.

First, at the initial stage, $t_\oplus\ll t_c$, viz., the second term
on the left-hand side of equation (17) can be neglected. The evolution of
the afterglow is the same as in the above case without any energy injection.

Second, for $T>t_\oplus\gg t_c$, the term $E_0$ in equation (17)
can be neglected. At this stage, the shock's velocity $v\propto
t_\oplus^{-2/5}$, its radius $r\propto t_\oplus^{3/5}$,
the internal field strength $B'\propto t_\oplus^{-2/5}$ and the typical
electron Lorentz factor $\gamma_{em}\propto t_\oplus^{-4/5}$. Thus,
we obtain the synchrotron peak frequency $\nu_m\propto \gamma_{em}^2B'
\propto t_\oplus^{-2}$, the cooling frequency $\nu_c\propto B'^{-3}
t_\oplus^{-2}\propto t_\oplus^{-4/5}$, the self-absorption frequency
$\nu_a\propto t_\oplus^{-2(p-1)/(p+4)}$, and the peak flux $F_{\nu_m}\propto
N_eP'_{\nu_m}\propto r^3B'\propto t_\oplus^{7/5}$. According to these
scaling laws, we derive the spectrum and light curve of the afterglow 
\begin{equation}
F_\nu=\left \{
       \begin{array}{lll}
         (\nu_a/\nu_m)^{-(p-1)/2}(\nu/\nu_a)^{5/2}F_{\nu_m}
         \propto \nu^{5/2}t_\oplus^{7/5} & {\rm if}\,\, \nu<\nu_a \\
         (\nu/\nu_m)^{-(p-1)/2}F_{\nu_m} \propto \nu^{-(p-1)/2}
         t_\oplus^{(12-5p)/5} & {\rm if}\,\, \nu_a< \nu <\nu_c \\
         (\nu_c/\nu_m)^{-(p-1)/2}(\nu/\nu_c)^{-p/2}F_{\nu_m}\propto
              \nu^{-p/2}t_\oplus^{2-p} & {\rm if}\,\, \nu>\nu_c.
        \end{array}
       \right.
\end{equation}
It can be seen that for high-frequency radiation the temporal decay index
$\alpha=(12-5p)/5\approx -0.4$ for emission from slow-cooling electrons
or $\alpha=2-p\approx -0.8$ for emission from fast-cooling electrons
if $p\approx 2.8$. This shows that the afterglow decay may significantly
flatten due to the effect of the pulsar.

Third, for $t_\oplus\gg T$, the power of the pulsar due to magnetic dipole
radiation rapidly decreases as $L\propto t_\oplus^{-2}$, and the evolution
of the shock is hardly affected by the stellar radiation. Thus,
the evolution of the afterglow at this stage will be the same as in
the above case without any energy injection.

\begin{center}
3. OBSERVED AND INFERRED PARAMETERS OF GRB 980519
\end{center}

We have shown that as an adiabatic shock expands in a dense medium
from an ultrarelativistic phase to a nonrelativistic phase, the decay
of radiation from such a shock will steepen, subsequently may flatten
if a strongly magnetic millisecond pulsar continuously inputs
its rotational energy into the shock through magnetic dipole radiation,
and finally the decay will steepen again due to disappearance of the stellar
effect. We next show that these effects can fit very well the observed
afterglow of GRB 980519.

GRB 980519 was one of the brightest of the bursts detected by the BeppoSAX
satellite (Muller et al. 1998; in 't Zand et al. 1999). The BATSE measured
fluence above 25 keV was $(2.54\pm 0.41)\times 10^{-5}\,{\rm ergs}\,{\rm cm}
^{-2}$, which places it among the top 12\% of BATSE bursts
(Connaughton 1998). An X-ray afterglow was detected by the BeppoSAX NFI
(Nicastro et al. 1999). The optical afterglow $\sim 8.5$ hours after the
burst presented the most rapid fading of the well-detected GRB afterglows
except for GRB 990510, consistent with $t_\oplus^{-2.05\pm 0.04}$ in
{\em BVRI} (Halpern et al. 1999), while the power-law decay index of the
X-ray afterglow, $\alpha_X=2.07\pm 0.11$ (Owens et al. 1998), in agreement
with the optical. The spectrum in optical band alone is well fitted by
a power low $\nu^{-1.20\pm 0.25}$, while the optical and X-ray spectra
together can also be fitted by a single power law of the form
$\nu^{-1.05\pm 0.10}$. In addition, the radio afterglow of this burst
was observed by the VLA at 8.3 GHz, and its temporal evolution
$\propto t_\oplus^{0.9\pm 0.3}$ between 1998 May 19.8UT and 22.3UT
(Frail, Taylor \& Kulkarni 1998).

We now analyze the observed afterglow data of GRB 980519 based on our
model. We assume that for this burst, the forward shock evolved from
an ultrarelativistic phase to a nonrelativistic phase in a dense
medium at $\sim 8$ hr after the burst. So, the detected afterglow,
in fact, was the radiation from a nonrelativistic shock. This implies
$\gamma\sim 1$ at $t_b\approx 1/3$ days. From equation (2), therefore,
we find
\begin{equation}
n_5\sim 27E_{52}[(1+z)/2]^3.
\end{equation}
If $p\approx 2.8$, and if the observed optical afterglow was emitted by
slow-cooling electrons and the X-ray afterglow from fast-cooling
electrons, then according to equation (16), the decay index $\alpha_R
=(21-15p)/10\approx -2.1$ and $\alpha_X=(4-3p)/2\approx -2.2$,
in excellent agreement with observations. Furthermore, the model spectral
index at the optical to X-ray band and the decay index at the
radio band, $\beta=-(p-1)/2 \approx -0.9$ and $\alpha=1.1$, are quite
consistent with the observed ones, $-1.05\pm 0.10$ and $0.9\pm 0.3$,
respectively.

We next continue to take into account three observational results. First,
on May 21.6UT, the Keck II 10m telescope detected the R-band magnitude
$R=23.03\pm 0.13$, corresponding to the flux $F_R\sim
3.5\mu{\rm Jy}$ at $t_\oplus\approx 2$ days (Gal et al. 1998).
Considering this result in the second sub-equation of (16) together with
equations (12), (15) and (19), we can derive
\begin{equation}
\epsilon_e^{1.8}\epsilon_{B,-6}^{0.95}E_{52}^{1.95}
\left(\frac{1+z}{2}\right)^{1.95}\left(\frac{\sqrt{1+z}-1}{\sqrt{2}-1}
\right)^{-2}\sim 1.7,
\end{equation}
where $p\approx 2.8$ has been assumed.

Second, the BeppoSAX observed the X-ray (2-10\,keV) flux $F_X\sim 1.3\times
10^{-2}\,\mu$Jy at $t_\oplus\approx 0.65$ days (Nicastro et al. 1999).
Inserting this result into the third sub-equation of (16) together with
equations (12), (13), (15) and (19), we can also derive
\begin{equation}
\epsilon_e^{1.8}\epsilon_{B,-6}^{0.2}E_{52}^{1.2}
\left(\frac{1+z}{2}\right)^{0.7}\left(\frac{\sqrt{1+z}-1}{\sqrt{2}-1}
\right)^{-2}\sim 0.14.
\end{equation}

Third, the VLA detected the radio flux $F_{\rm 8.3 GHz}\approx 102\pm 19\,
\mu{\rm Jy}$ on May 22.3UT in 1998 (Frail et al. 1998). This may result from
the self-absorption effect in the shocked medium. Thus, combining it with
the first sub-equation of (16) together with equations (12), (14), (15)
and (19) leads to
\begin{equation}
\epsilon_e^{0.32}\epsilon_{B,-6}^{-0.24}E_{52}^{-1.28}
\left(\frac{1+z}{2}\right)^{-0.26}\left(\frac{\sqrt{1+z}-1}{\sqrt{2}-1}
\right)^{-2}\sim 3.2.
\end{equation}

In addition, the total energy of the adiabatic shock, $E_0$, is
approximately equal to the one released initially in gamma-rays (Piran 1999).
This implies
\begin{equation}
E_{52}\sim 10\left(\frac{1+z}{2}\right)\left(\frac{\sqrt{1+z}-1}{\sqrt{2}-1}
\right)^2.
\end{equation}

From equations (19)-(23), we infer intrinsic parameters of the shock and
the redshift of the burst:
\begin{equation}
  \begin{array}{lll}
  \epsilon_e\sim 0.16,\,\,\,\epsilon_{B,-6}\sim 280, \\
  E_{52}\sim 0.27,\,\,\,n_5\sim 3.4, \\
  z\sim 0.55.
  \end{array}
\end{equation}
Our inferred value of $\epsilon_e$ is near the equipartition value,
in agreement with the result of Wijers \& Galama (1998) and Granot,
Piran \& Sari (1998), while our $\epsilon_B$ is also close to the value
inferred from GRB 971214 and GRB 990123 (Wijers \& Galama 1998;
Galama et al. 1999; Wang et al.  1999a). After considering these reasonable
parameters, Wang, Dai \& Lu (1999b) numerically studied the trans-relativistic
evolution of the shock and found that our dense medium model can provide
an excellent fit to all the observational data of the radio afterglow from 
GRB 980519 shown in Frail et al. (1999). 

If the late afterglow of GRB 980519 had still decayed according to equation
(16), the inferred R-band fluxes on the 60th day and 66th day would have
been nearly two orders of magnitude smaller than the observed values.
This would lead to the argument that the emission on these two days came
from the host galaxy of the burst (Sokolov et al. 1998; Bloom et al. 1998).
We note that, despite of excellent seeing conditions on the Keck II telescope,
Bloom et al. (1998) found little evidence for extension expected of
a host galaxy. This implies that there may exist some mechanism by which
the shock at the late stage had been renewed. As suggested in the above
section, this mechanism is that a strongly magnetized millisecond
pulsar had supplied its rotational energy to the shock through magnetic
dipole radiation. We can see from the second sub-equation of (18) that when
$t_c\ll t_\oplus< T$, the R-band afterglow decay index
$\alpha_R=(12-5p)/5\approx -0.4$, where $p\approx 2.8$ has been assumed.
Combining this result with the observed flux $F_R\sim 0.2\,\mu{\rm Jy}$
on the 60th day and the decay power law in several days after the burst,
we infer $t_c\sim 4$ days. According to the definition of
$t_c=29E_{52}B_{s,13}^{-2}P_{\rm ms}^4\,$days where $B_{s,13}$ is the surface
magnetic field strength of the pulsar in units of $10^{13}\,$G and
$P_{\rm ms}$ is its initial period in units of 1 ms (Dai \& Lu 1998a, c),
we can obtain a constraint on the stellar parameters:
$B_{s,13}\sim 2.7E_{52}^{1/2}P_{\rm ms}^2\sim 1.7P_{\rm ms}^2$.
Moreover, our model requires $T> 66$ days, which leads to $P_{\rm ms}<0.8$,
where we have used the definition of the stellar spindown timescale,
$T=120B_{s,13}^{-2}P_{\rm ms}^2\,$days. Therefore, if GRB 980519 resulted
from a pulsar, and if the property of the late afterglow was caused by
the effect of the stellar magnetic dipole radiation, then this pulsar may
be a strongly magnetized millisecond or even submillisecond one.

\begin{center}
4. PROPERTIES OF OTHER BURSTS
\end{center}

\begin{center}
4.1. {\em GRB 990510}
\end{center}

GRB 990510 was detected by the BeppoSAX Gamma-Ray Burst Monitor
(Piro et al. 1999) as a bright and complex GRB composed by two well
seperated and multi-peaked pulses with a total duration of about 75 s
(Amati et al. 1999). Its fluence was among the highest of the BeppoSAX
localized events, after GRB 990123, GRB 980329 and GRB 970111.
It was also detected by BATSE (Kippen et al. 1999) and
its fluence ($> 20$ keV) was $(2.56\pm 0.09) \times 10^{-5}\,{\rm erg}\,
{\rm cm}^{-2}$, in the top 9\% of the BATSE burst fluence distribution.
The burst appears at $z\ge 1.62$ (Vreeswijk et al. 1999),
which leads to an isotropic energy of $\ge 1.4\times 10^{53}\,
{\rm ergs}$. The burst's afterglow was detected and
monitored at X-ray and optical bands. Even though the X-ray afterglow
decay light curve is not unlike that seen previous X-ray afterglow decays
(Kuulkers et al. 1999), the optical afterglow displays its special feature:
all the temporal decays at {\em VRI} bands steepened about 1.2 days after
the burst (Harrison et al. 1999; Stanek et al. 1999; Bloom et al. 1999a;
Marconi et al. 1999). Initially the optical decay index
$\alpha_1=-0.82\pm 0.02$, but about 1.2 days later the index
became $\alpha_2=-2.18\pm 0.05$. The consistency of $\alpha_1$, $\alpha_2$
and the breaking time means that the breaking is wide band. This is
the first clear observation of a wide band break (Bloom et al. 1999a;
Harrison et al. 1999).

One simple interpretation for this steepening seems that we have been seeing 
evidence for a spreading jet (Rhoads 1999; Sari et al. 1999; 
Bloom et al. 1999a). As shown numerically in Panaitescu \& M\'esz\'aros 
(1999), Moderski et al. (1999) and Huang et al. (1999b, 2000), however,
the evolution of a spreading jet may not lead to a marked steepening.
Another possible interpretation is that the effect of a strongly magetized
millisecond pulsar on the evolution of a nonrelativistic adiabatic shock
in a dense medium has been becoming unimportant. If initially the pulsar
was able to change the evolution of the shock for GRB 990510 through
magnetic dipole radiation, and if the optical afterglow
came from fast-cooling electrons in the shocked medium, then according to
the second sub-equation of (18), the temporal decay index $\alpha=2-p\approx
-0.82\pm 0.02$. This requires $p\approx 2.8$, which is quite consistent
with the value inferred from GRB 980519. When the effect of the pulsar
on the shock disappeared, the optical afterglow decayed based on the third
sub-equation of (16), viz., $\alpha=(4-3p)/2\approx -2.2$, in excellent
agreement with the observations. Furthermore, the observed breaking time
should be equal to the stellar spindown timescale in our model, which 
constrains the pulsar's field strength: $B_{s,13}\sim 10P_{\rm ms}$.
Thus, the central engine of this burst could be a millisecond magnetar.

\begin{center}
4.2. {\em GRB 980326}
\end{center}

The afterglow of GRB 980326 also had a rapid decline. Groot et al. (1998)
derived a temporal decay index of $\alpha=-2.1\pm 0.13$
and a spectral index of $\beta=-0.66\pm 0.7$ in the optical band.
This initial decay index, which is similar to that of GRB 980519,
suggests the evolution of a nonrelativistic adiabatic shock in a dense
medium. There is another observational result similar to the case of
GRB 980519: the decay of the observed optical afterglow began to flatten
about 5 days after the burst; this is not the contribution of the host
galaxy because it is not present at a later time (Bloom \& Kulkarni 1998;
Bloom et al. 1999b). Consequently, the late afterglow might be interpreted
as a different phenomenon. Bloom et al. (1999b) suggested that
this late afterglow could result from a supernova associating with
GRB 980326. In our model, this is understood to be the emission from the
nonrelativistic shock to which has been input energy by a strongly magnetized
millisecond pulsar. According to the light curve shown in Figure 2 of
Bloom et al. (1999b), we infer $t_c\sim 5$ days. Thus, the surface
field strength of the pulsar could be $B_{s,13}\sim 2.4E_{52}^{1/2}
P_{\rm ms}^2$.

\begin{center}
5. DISCUSSIONS AND CONCLUSIONS
\end{center}

There have been two kinds of plausible models for GRB energy sources
in the literature: one relating to pulsars and another to stellar-mass
black holes. Dai \& Lu (1998c) have summarized possible progenitors
involving strongly magnetic millisecond pulsars: accretion-induced
collapses of magnetized white dwarfs, mergers of two neutron stars if
the equation of state for neutron matter is moderately stiff to stiff,
and phase transitions of neutron stars to strange stars. The pulsar
progenitor 
models also include the birth of magnetars in supernova explosions
suggested by Wheeler et al. (1999). The rotational
energy of such pulsars at the moment of their formation is as high as a few
$\times 10^{53}$ ergs. The efficiency of transformation of the rotational
energy to the energy of a relativistic outflow and then to the energy
of high-frequency radiation may be as high as almost 100\% (Usov 1994;
Blackman, Yi \& Field 1996). Such pulsars have been suggested to generate
possibly anisotropic outflows (Dai \& Lu 1998a; Smolsky \& Usov 1999),
and thus may explain the energetics of GRBs, including GRB 990123
as an extreme event if the energy flux from the source at the line of sight
is only about ten times more than the energy flux averaged over
all directions. Two by-products in this kind of models are millisecond
pulsars and relativistic forward shocks forming during the collision of
the outflows with ambient media. It is natural to expect that the central
pulsars affect the evolution of postburst shocks and in turn the afterglows
from these shocks through magnetic dipole radiation.

In the second kind of GRB source models, Fryer, Woosley \& Hartmann (1999)
have summarized possible progenitors involving black hole accretion disks:
neutron star-neutron star binary mergers, black hole-neutron star mergers,
black hole-white dwarf mergers, massive star core collapses,
and black hole-helium star mergers. This kind of models should include
the supranovae proposed by Vietri \& Stella (1998).

Some of the source models mentioned above, e.g., failed supernovae
(Woosley 1993), hypernovae (Paczy\'nski 1998), supranovae (Vietri \&
Stella 1998), phase transitions of neutron stars to strange stars
(Dai \& Lu 1998a; DL99) and birth of magnetars (Wheeler et al. 1999), 
may lead to dense media prior to the occurrence of GRBs. 
In addition, dense media have also been discussed in the context
of GRBs. For example, Katz (1994) suggested collisions of relativistic
nucleons with a dense cloud as an explanation of the delayed hard
photons from GRB 940217. DL99 discussed the evolution of an adiabatic
shock in a dense medium to explain the steepening feature in the light
curve of the R-band afterglow from GRB 990123.

Based on these arguments, following DL99, we discuss the evolution of an
adiabatic shock expanding in a dense medium from an ultrarelativistic phase
to a nonrelativistic phase in more details in this paper. In particular,
we discuss the effects of synchrotron self absorption and energy
injection on the afterglow from this shock. In a dense medium, the shock
becomes nonrelativistic rapidly after a short relativistic phase.
This transition time varies from several hours to a few days when
the medium density is from $10^5$ to a few $\times 10^6\,{\rm cm}^{-3}$,
and the shock energy from $10^{51}$ to $10^{54}$ ergs.
The afterglow from the shock at the nonrelativistic stage 
decays more rapidly than at the relativistic stage, while the decay
index varies from $-1.35$ to $-2.1$ if the spectral index of
the accelerated electron distribution, $p=2.8$, and
the radiation comes from those slow-cooling electrons. Since some models
mentioned above predict that a strongly magnetic millisecond pulsar
may be born during the formation of GRB, we also discuss the effect of such
a pulsar on the evolution of a nonrelativistic shock through magnetic dipole
radiation, in contrast to the case discussed in Dai \& Lu (1998a, c).
We find that after the energy which the shock obtains from the pulsar
is much more than the initial energy of the shock, the afterglow decay
will flatten significantly and the decay index will become $-0.4$.
When the pulsar energy input effect disappears, the index will still be
$-2.1$.
These features are in excellent agreement with the afterglows of
GRB 980519 and GRB 980326. Furthermore, our model fits very well
all the observational data of GRB 980519 including the last two detections.
Of course, if an afterglow of our interest comes from fast-cooling
electrons in the shocked medium, the decay index of this afterglow will
first be $-1.6$ during the relativistic phase and subsequently $-2.2$ at the
nonrelativistic stage in the case of $p=2.8$. If the energy input effect 
of the pulsar during this stage becomes very important, the decay index 
will be $-0.8$. When this effect disappears, the index will become
$-2.2$ again. The latter values of the index are quite consistent with
the observations of GRB 990510. This requires that the time before
the pulsar is able to affect the evolution of the shock is only as short
as a few hours. It should be pointed out that whether and when a pulsar
significantly affects the evolution of a shock largely depends upon
the following three parameters: the shock's initial energy, the dipole
magnetic field and period of the pulsar.

The flattening of the late optical afterglow light curves of GRB 980519 
and GRB 980326 has also been interpreted as being due to Ib/c supernovae, 
which is based on the obvious reddening of the observed afterglow 
spectrum. In our model, this flattening is due to the energy input effect of 
millisecond pulsars through magnetic dipole radiation, and the expected 
afterglow spectrum would be the typical synchrotron spectrum without 
any dust effect. If dust exists in the vicinity of the pulsars, however, dust 
sublimation may lead to reddening of the afterglow spectrum as suggested 
by Waxman \& Draine (1999). In explaining the steepening of the afterglow 
light curve of GRB 990510, we have envisioned the disappearance of the 
pulsar energy input effect. In our model,  radio afterglows first brighten as 
$\propto t_\oplus^{5/4}$ with the pulsar energy input effect, and 
subsequently fade down as $\propto t_\oplus^{(4-3p)/2}$ when this effect 
becomes unimportant and the observed frequency is smaller than the
synchrotron 
self-absorption frequency. Such a steepening has also been understood to 
be due to the lateral expansion of relativistic jets (Rhoads 1999; Sari et
al. 1999). 
Thus, radio afterglows from jets first decay as $\propto t_\oplus^{-1/3}$ and 
then as $\propto t_\oplus^{-p}$  (Sari et al. 1999). Therefore, there is an
obvious 
difference in radio afterglow light curves between the pulsar energy input 
model and the jet model. We expect that future observations will provide 
evidence for or against the pulsar energy input model.   

In summary, following DL99, we propose a model for several afterglows,
in which a shock expanding in a dense medium evolves if its central engine
is a strongly magnetized millisecond pulsar. We show that this model
explains well the features of the afterglows from GRBs 980519, 980326
and 990510.

We would like to thank the anonymous referee for valuable comments 
and suggestions, and Y. F. Huang and D. M. Wei for helpful discussions.
This work was supported partially by the National Natural Science 
Foundation of China (grants 19825109 and 19773007) and partially by 
the National 973 Project.

\newpage 
\baselineskip=2.5mm

\begin{center}
REFERENCES
\end{center}

\begin{description}
\item Amati, L., Frontera, F., Costa, E., \& Feroci, M. 1999, GCN Circ. 317
\item Blackman, E. G., Yi, I., \& Field, G. B. 1996, ApJ, 473, L79
\item Blandford, R. D., \& McKee, C. F. 1976, Phys. Fluids, 19, 1130
\item Bloom, J. S., \& Kulkarni, S. R. 1998, GCN Circ. 161
\item Bloom, J. S., Kulkarni, S. R., Djorgovski, S. G., Gal, R. R.,
          Eichelberger, A., \& Frail, D. A. 1998, GCN Circ. 149
\item Bloom, J. S. et al. 1999a, GCN Circ. 323
\item Bloom, J. S. et al. 1999b, Nature, 401, 453
\item Castro-Tirado, A. J. et al. 1999, Science, 283, 2069
\item Chiang, J., \& Dermer, C. D. 1998, astro-ph/9803339
\item Chevalier, R. A., \& Li, Z. Y. 1999a, ApJ, 520, L29
\item Chevalier, R. A., \& Li, Z. Y. 1999b, astro-ph/9908272
\item Cohen, E., Piran, T., \& Sari, R. 1998, ApJ, 509, 717
\item Connaughton, V. 1998, GCN Circ. 86
\item Dai, Z. G., Huang, Y. F., \& Lu, T. 1999, ApJ, 520, 634
\item Dai, Z. G., \& Lu, T. 1998a, Phys. Rev. Lett., 81, 4301
\item Dai, Z. G., \& Lu, T. 1998b, MNRAS, 298, 87
\item Dai, Z. G., \& Lu, T. 1998c, A\&A, 333, L87
\item Dai, Z. G., \& Lu, T. 1999, ApJ, 519, L155 (DL99)
\item Djorgovski, S. G. et al. 1997, Nature, 387, 876
\item Frail, D. A., Taylor, G. B., \& Kulkarni, S. R. 1998, GCN Circ. 89
\item Frail, D. A., et al. 1999, ApJL, submitted (astro-ph/9910060)
\item Fruchter, A. S. et al. 1999, ApJ, in press (astro-ph/9902236)
\item Fryer, C. L., Woosley, S. E., \& Hartmann, D. H. 1999, ApJ, submitted \\
          (astro-ph/9904122)
\item Gal, R. R. et al. 1998, GCN Circ. 88
\item Galama, T. J. et al. 1999, Nature, 398, 394
\item Granot, J., Piran, T., \& Sari, R. 1998, astro-ph/9808007
\item Groot, P. J. et al. 1998, ApJ, 502, L123
\item Halpern, J. P., Kemp, J., Piran, T., \& Bershady, M. A. 1999,
          ApJ, submitted  \\ (astro-ph/9903418)
\item Harrison, F. A. et al. 1999, astro-ph/9905306
\item Huang, Y. F., Dai, Z. G., \& Lu, T. 1998, A\&A, 336, L69
\item Huang, Y. F., Dai, Z. G., \& Lu, T. 1999a, MNRAS, 309, 513
\item Huang, Y. F., Dai, Z. G., Wei, D. M., \& Lu, T. 1998, MNRAS, 298, 459
\item Huang, Y. F., Gou, L. J., Dai, Z. G., \& Lu, T. 1999b, ApJ, submitted
          (astro-ph/9910493)
\item Huang, Y. F., Dai, Z. G., \& Lu, T. 2000, MNRAS, submitted
\item in 't Zand, J. J. M., Heise, J. van Paradijs, J., \& Fenimore, E. E.
          1998, ApJ, in press
\item Katz, J. I. 1994, ApJ, 432, L27
\item Katz, J. I., \& Piran, T. 1997, ApJ, 490, 772
\item Kippen, R. M. 1999, GCN Circ. 322
\item Kulkarni, S. R. et al. 1999, Nature, 398, 389
\item Kuulkers, E. et al. 1999, GCN Circ. 326
\item Marconi, G., Israel, G. L., Lazzati, D., Covino, S., \&
          Ghisellini, G. 1999, GCN Circ. 329
\item M\'esz\'aros, P., \& Rees, M. J. 1997, ApJ, 476, 232
\item M\'esz\'aros, P., Rees, M. J., \& Wijers, R. A. M. J. 1998,
          ApJ, 499, 301
\item Moderski, R., Sikora, M., \& Bulik, T. 1999, astro-ph/9904310
\item Muller, J. M. et al. 1998, IAU Circ. 6910
\item Nicastro, L. et al. 1999, A\&A, in press (astro-ph/9904169)
\item Owens, A. et al. 1998, A\&A, 339, L37
\item Paczy\'nski, B. 1998, ApJ, 494, L45
\item Panaitescu, A., \& M\'esz\'aros, P. 1999, ApJ, in press
          (astro-ph/9806016)
\item Panaitescu, A., M\'esz\'aros, P., \& Rees, M. J. 1998, ApJ, 503, 315
\item Piran, T. 1999, Phys. Rep., 314, 575
\item Piro, L. et al. 1999, GCN Circ. 304
\item Rees, M. J., \& M\'esz\'aros, P. 1998, ApJ, 496, L1
\item Reichart, D. E. 1997, ApJ, 485, L57
\item Rhoads, J. 1999, ApJ, in press (astro-ph/9903399)
\item Rybicki, G. B. \& Lightman, A. P. 1979, Radiative Processes
          in Astrophysics (New York: Wiley)
\item Sari, R. 1997, ApJ, 489, L37
\item Sari, R., \& Piran, T., \& Halpern, J. P. 1999, astro-ph/9903339
\item Sari, R., Piran, T., \& Narayan, R. 1998, ApJ, 497, L17
\item Smolsky, M. V., \& Usov, V. V. 1999, astro-ph/9905142
\item Sokolov, V., Zharikov, S., Palazzi, L., \& Nicastro, L. 1998,
          GCN Circ.148
\item Stanek, K. Z. et al. 1999, astro-ph/9905304
\item Usov, V. V. 1994, MNRAS, 267, 1035
\item Vietri, M. 1997, ApJ, 488, L105
\item Vietri, M., \& Stella, L. 1998, ApJ, 507, L45
\item Vreeswijk, P. M. et al. 1999, GCN Circ. 324
\item Wang, X. Y., Dai, Z. G., \& Lu, T. 1999a, MNRAS, submitted
          (astro-ph/9906062)
\item Wang, X. Y., Dai, Z. G., \& Lu, T. 1999b, MNRAS, submitted
          (astro-ph/9912492)
\item Waxman, E. 1997a, ApJ, 485, L5
\item Waxman, E. 1997b, ApJ, 489, L33
\item Waxman, E., \& Draine, B. T. 1999, ApJ, submitted (astro-ph/9909020)
\item Waxman, E., Kulkarni, S. R., \& Frail, D. A. 1998, ApJ, 497, 288
\item Wheeler, J. C., Yi, I., H\"oflich, P., \& Wang, L. 1999, ApJ, in press
\item Wijers, R. A. M. J., \& Galama, T. J. 1998, ApJ, in press
          (astro-ph/9805341)
\item Wijers, R. A. M. J., Rees, M. J., \& M\'esz\'aros, P. 1997, MNRAS,
          288, L51
\item Woosley, S. 1993, ApJ, 405, 273
\item Zharikov, S. V., Sokolov, V. V., \& Baryshev, Yu V. 1998, A\&A, 337, 356
\end{description}

\end{document}